**Reducing critical current for spin-transfer-torque-induced magnetization reversal in current-perpendicular-to-plane giant magnetoresistance devices: effect of low damping and enhanced spin scattering asymmetry in $Co_2FeGa_{0.5}Ge_{0.5}$ Heusler alloy**

Vineet Barwal, Hirofumi Suto*, and Yuya Sakuraba

Research Center for Magnetic and Spintronic Materials, National Institute for Materials Science (NIMS), Tsukuba, 305-0047, Japan

*SUTO.Hirofumi@nims.go.jp

**Abstract**

Spin-transfer torque (STT) in magnetoresistance devices has enabled key applications such as STT-magnetoresistive random access memory, spin torque oscillators, and energy-assisted magnetic recording. In the device structures, where a free layer (FL) magnetization is manipulated by spin injection from a spin injection layer (SIL), the critical current density required for operation is directly proportional to the damping ($\alpha$) constant of FL and inversely proportional to the STT efficiency, which depends on the spin polarization ($P$) of the materials. Here, we investigate the effect of low $\alpha$ and high $P$ of $Co_2FeGa_{0.5}Ge_{0.5}$ (CFGG) Heusler alloy on the operation current required for STT-induced magnetization reversal in current perpendicular-to-plane giant magnetoresistance devices. Devices with CFGG as a FL material achieved a large reduction in the operation current, as compared to those with conventional NiFe-FL owing to the very low $\alpha$ of CFGG, demonstrating the advantage of CFGG as a FL material. As the advantage of high spin polarization CFGG for SIL, we analyzed the effect of bilayer SIL consisting of CoFe and thin CFGG layers, focusing on utilizing the spin scattering asymmetry at the CoFe/CFGG interface. Devices with the CoFe/CFGG-SIL exhibited the lowest critical current, demonstrating enhanced STT efficiency. In addition, the correlation of STT efficiency with magnetoresistance ratio were comprehensively investigated, showing that device-to-device distribution in STT-efficiency was smaller in CoFe/CFGG-SIL. These findings highlight the potential of CFGG Heusler alloy and CoFe/CFGG bilayer structures as key components for the development of efficient and stable STT-based spintronic devices.

1. Introduction

Spin transfer torque (STT) has been widely used to manipulate the magnetization direction of ferromagnets in various spintronic devices such as spin-torque oscillators[1,2], magnetoresistive random-access memory[3,4], and emerging computing devices[5]. In particular, STT realized in current perpendicular-to-plane giant magnetoresistance (CPP-GMR) devices have been extensively studied for assisted writing in hard disk drives (HDDs) to enhance the areal density[6,7,8,9,10,11]. In this application, STT-based devices are fabricated in the write gap of the write head, and induced magnetization excitation enhances the writability through the flux control (FC) effect[12,13] or the microwave assistance effect[14,15]. Here, we explain the operation principle of FC devices as it is closely related to this study. The simplest FC device has a tri-layer structure consisting of a magnetic free layer (FL), a metallic non-magnetic spacer layer and a magnetic spin injection layer (SIL). SIL is in direct contact and magnetically coupled with the pole materials making the SIL magnetization stable against STT. When a sufficient bias current is applied, STT via spin injection from SIL induce magnetization reversal of FL against the field inside the write gap (gap field). The dipolar field from the reversed FL magnetization can enhance the amplitude and gradient of the write field. The reliable and fast operation of the FC devices has already been demonstrated using actual HDDs[12,13,14].

One of the primary objectives in these STT-based device applications is to increase the efficiency of magnetization manipulation. In assisted writing, higher efficiency can induce excitation of a larger magnetic volume within the current tolerance of the devices, leading to larger assistance effect. Similarly, in memory and computation applications, higher efficiency reduces the operational current density, thereby lowering power consumption. According to Slonczewski's theory[16], the critical current density of STT-driven magnetization dynamics in CPP-GMR type tri-layer structure is expressed as:

$$J_\mathrm{c} = \mu_0 \frac{2|e|}{\hbar \eta} \alpha M_s^\mathrm{FL} d H_\mathrm{eff}. \qquad (1)$$

where $d$, $M_s^\mathrm{FL}$ and $\alpha$ is the thickness, saturation magnetization and Gilbert damping parameter of FL, respectively. STT efficiency, $\eta$ is a dimensionless parameter that depends on the spin polarization ($P$) of the conduction electron in the magnetic layers and the relative angle between magnetization directions of the FL and SIL. $H_{\text{eff}}$ is the effective field acting on FL. To reduce the operational current density for STT-induced magnetization manipulation, FL material with low $\alpha$ and SIL materials with high $P$ are beneficial because $J_c$ is directly proportional to $\alpha$ and high $P$ enhances $\eta$.

Furthermore, in assisted writing applications, the SIL magnetization counteracts the assistance effect by degrading the recording field. Therefore, reducing the magnetic volume of the SIL is crucial for enhancing the device performance. Thin SIL is also desirable because the total thickness of the device structure need to fit in small write gap of typically ~ 20 nm[14].

To meet the above requirements, Co-based half-metallic ferromagnet Heusler alloys are promising candidates for use as magnetic layers in STT-based devices due to their theoretically predicted large $P$ (~ 100%), low $\alpha$ (~ $1 \times 10^{-4}$) and high Curie temperature (> 600 K)[3,17,18,19]. In addition, the Heusler alloys exhibit relatively short spin-diffusion length ($\lambda_{SDL}$ ~ 2-6 nm) which enable them to function effectively as thin layers[20,21,22]. Among these materials, Co$_2$FeGa$_{0.5}$Ge$_{0.5}$ (CFGG) is selected in this study because of the reported low $\alpha$[23] and theoretically predicted half-metallicity[24,25]. Several studies have reported large magnetoresistance (MR) ratios in CPP-GMR devices incorporating CFGG, demonstrating its potential for enhancing spintronic device performance[22,24,26,27,28,29,30,31,32,33]. The observed large MR ratios are attributed to the high $P$ of CFGG. However, its impact on STT efficiency has not been studied in detail.

In this study, we first performed a detailed structural analysis and evaluated the $\alpha$ of single-layer CFGG Heusler thin films annealed at various temperatures. Subsequently, the CFGG Heusler was incorporated in CPP-GMR devices to investigate the influence of its high $P$ and low $\alpha$ on the MR and STT properties. The significantly lower damping parameter of CFGG compared to conventional NiFe (Py), led to reduction in the critical current required for STT-induced magnetization reversal against an applied magnetic field, demonstrating its potential for efficient STT-based device operation. Furthermore, to leverage the high spin polarization of CFGG for spin injection, we investigated a bilayer SIL composed of CoFe and a thin CFGG layer. The idea of using CoFe/thin CFGG bilayer as SIL is derived from the following two facts: the device structures proposed for assisted writing use SIL attached to CoFe-based magnetic poles[13] and the enhancement in the MR ratio is reported by using the CoFe/CFGG bilayer due to the spin scattering asymmetry at the interface[34]. The use of CoFe/CFGG bilayers as SIL led to a further reduction in operation current with reduced device-to-device distribution. The stable operation and enhanced efficiency achieved with CoFe/CFGG-SIL configurations further underscore their suitability for STT-based devices. These findings highlight the advantages of CFGG as a low-damping FL and CoFe/CFGG as a stable and efficient SIL, making CFGG a promising candidate material for high-performance STT-based devices.

## 2. Experimental Details

**Sample growth and Device Fabrication**

We prepared the following three series of samples by magnetron sputtering on MgO(001) substrates: (A) single-layer CFGG films for structural analysis and $\alpha$ evaluation by ferromagnetic resonance (FMR) measurement, (B) GMR stacks with CFGG and Py as FL, and (C) GMR stacks with CoFe/CFGG bilayers as SIL. The order of the stacking structure is from bottom to top and the numbers in the parentheses represent the nominal thickness in nm. The series-A samples composed of CFGG(30)/Al(2) films. These films were subjected to in-situ annealing at post-annealing temperature ($T_p$) of 300, 400, 500 and 600 °C. The Al capping layer was deposited after cooling down the sample to room temperature (RT). Additionally, two control samples with stacking structures Cr(5)/Ag(5)/CFGG(6)/Ru(8) and Cr(5)/Ag(5)/Py(7)/Ru(8) were prepared. The structures were designed to match the FL configuration in the series-B samples enabling a quantitative estimation of $\alpha$ for the CFGG and Py layers in the CPP-GMR stacks. The CFGG control sample was annealed at $T_p$ = 500 °C and cooled down to RT prior to deposition of the Ru layer.

In the series-B and -C samples, Cr(5)/Ag(100) bottom electrode was first deposited and annealed at 300 °C. After cooling the sample down to RT, the GMR stack was subsequently deposited. The series-B samples contain GMR stacks with two different FL configuration: CFGG(15)/Ag(7)/CFGG(6) annealed at $T_p$ = 500 °C after the top CFGG layer deposition and CFGG(15)/Ag(7)/Py(7.5) annealed at $T_p$ = 500 °C after bottom CFGG layer deposition. These samples are named as CFGG-FL and Py-FL, respectively. The bottom 15 nm CFGG layer act as the SIL, and its higher thickness prevents reversal of the SIL magnetization due to the counter spin injection from FL. Different FL thickness was chosen for CFGG-FL and Py-FL to keep the FL magnetization volume almost the same, considering the different saturation magnetization of CFGG and Py. The GMR stack of series-C samples composed of CoFe(6)/CFGG(3)/Ag(7)/CFGG(6) and were annealed at $T_p$ = 350, 400, 450 and 500 °C after the deposition of the top CFGG layer. These are referred to as CoFe/CFGG-SIL samples. The choice of the CoFe(6)/CFGG(3) configuration for the SIL is based on two key considerations. First, to effectively utilize the interfacial spin polarization at the CoFe/CFGG interface, the thickness of the CFGG layer was kept comparable to $\lambda_{SDL}$ (typically ~3 nm), allowing spin polarized current to be injected with minimal relaxation[22,27,34,35]. Second, the SIL must maintain a stable magnetization against reverse STT exerted by the 6 nm-thick CFGG FL. To achieve this, a 6 nm-thick CoFe layer is used in the SIL, providing sufficient magnetic volume to ensure magnetic stability.

The annealing time for the series-A and series-B samples was set to 30 minutes. In contrast, the annealing duration was reduced to 5 minutes for the series-C sample, to minimize interfacial interdiffusion between CoFe and CFGG layers. A similar short duration annealing was employed in Ref[36], where CPP-GMR structures containing CoFe/Co$_2$Fe$_{0.4}$Mn$_{0.6}$Si bilayer electrodes were annealed for 2 minutes. The annealing temperature in the series-B and the series-C samples was limited to 500°C as higher temperatures are generally known to cause interdiffusion and degrade the MR output[27,29,30]. The CFGG films were deposited using a composite sputtering target having the composition of Co$_{43.28}$Fe$_{26}$Ga$_{15.16}$Ge$_{15.56}$. The typical composition of the deposited CFGG layer was Co$_{45.20}$Fe$_{31.90}$Ga$_{12.69}$Ge$_{10.21}$, as measured by X-ray fluorescence analysis, calibrated by the standard sample whose composition was analyzed by inductively coupled plasma mass spectrometry. The Co-deficient and Fe-rich off-stoichiometric CFGG composition was chosen to suppress Co$_{Fe}$ antisites detrimental to $P$ while obtaining single-phase CFGG Heusler films[25,37,38].

Pseudo-spin-valve CPP-GMR devices were fabricated for the series-B and series-C stacks by the following procedure. The samples were patterned into circular and elliptical pillars with lateral dimensions of 200 × 200 nm$^2$, 200 × 100 nm$^2$, 100 × 100 nm$^2$ and 80 × 80 nm$^2$ dimensions by electron beam lithography and Ar ion milling. Then, the pillars were covered by a thin Ta adhesion layer and a SiO$_2$ passivation layer. After the lift-off of the covering layers, an Au top electrode was fabricated. For each device size, more than 50 devices were prepared on one substrate. The MR measurement was conducted on all the devices, and defective devices exhibiting abnormal resistance were excluded from the analysis. The STT measurement was conducted on 80 × 80 nm$^2$ devices. Since all devices used for STT measurements have identical lateral dimensions, the switching current $I_c$ is reported instead of current density, allowing for direct comparison without normalization. The circular pillar's cross-sectional area was estimated to be ~10.58 × 10$^{-3}$ μm$^2$ using scanning electron microscopy.

**Characterization**

The structural characterization for the series-A, series-B and series-C sample stacks was done using X-ray diffraction (XRD) with a Cu–K$_\alpha$ radiation source ($\lambda$ = 1.5406 Å). The magnetization dynamics properties for the series-A samples were measured by in-plane FMR measurements using broadband FMR set-up in the Physical Property Measurement System (PPMS Dynacool; Quantum Design). The in-plane resonance field ($\mu_0 H_r$) and linewidth ($\mu_0 \Delta H$) were obtained by fitting the FMR spectra to a Lorentzian derivative function containing both symmetric and anti-symmetric components, as detailed in Ref[39]. The damping parameter ($\alpha$) was then extracted using the following equation.

$$\mu_0 \Delta H = \mu_0 \Delta H_0 + \frac{4\pi\alpha f}{\gamma} \qquad (2)$$

where $\mu_0 \Delta H_0$ accounts for line-broadening owing to the extrinsic contributions such as magnetic inhomogeneities, $f$ is the frequency in the GHz range, $\gamma \left(= g \frac{\mu_B}{\hbar}\right)$ is the gyromagnetic ratio with $g \sim 2.1$ as the Lande's factor, $\mu_B$ as the Bohr magneton and $\hbar$ as the reduced Planck's constant.

Resistance versus magnetic field (*R-H*) measurements were conducted on the CPP-GMR devices fabricated from the series-B and series-C samples through the four-probe measurement method. The measurement was done in an auto-prober system by applying in-plane magnetic field along the long axis of the elliptical pillars. MR ratio was defined as ($R_P - R_{AP}$)/$R_P$ × 100%, where $R_P$ and $R_{AP}$ are the resistance in the parallel (P) and anti-parallel (AP) configuration of the FL and SIL. The STT induced magnetization reversal against the magnetic field measurements were conducted on the CPP-GMR devices from the series-B and series-C samples having circular pillars with designed diameter of 80 nm. The details of the measurement method can be found in the Ref[40,41].

## 3. Results and Discussion

### I. Series-A single-layer CFGG films

**Structural characterization**

XRD measurements were performed to study the effect of annealing temperature on the structural properties of the series-A single layer CFGG films. Figure 1 (a) shows the out-of-plane XRD profiles at $\chi = 0°$ for the series-A samples. The XRD profiles show only the diffraction peaks corresponding to the (001) plane, indicating (001)-oriented growth. The 002 Heusler peak confirms $B2$ ordering, reflecting the atomic order between Co and (Fe, Ga, Ge) sites in all the samples. The increasing intensity of 002 peak with higher $T_p$, suggests improved $B2$ ordering. Figure 1(b) shows the corresponding XRD profiles along the [111] direction at $\chi = 54.7°$. A faint 111 superlattice diffraction peak is observed for $T_p$ = 500 °C sample and becoming more prominent at $T_p$ = 600 °C, indicating the enhancement of $L2_1$ ordering, which reflects atomic ordering between Fe, and (Ga, Ge sites). The lattice mismatch between CFGG and MgO is estimated to be < 3.7 % from the lattice parameter, $a$ = 0.574 nm for the CFGG and $\sqrt{2}a$ = 0.595 nm for MgO. The lattice parameter matches well with the reported bulk value, exhibiting values within the range of 0.572-0.573 nm with different $T_p$. The measured XRD profiles for the control samples are shown in the supplementary figure S1.

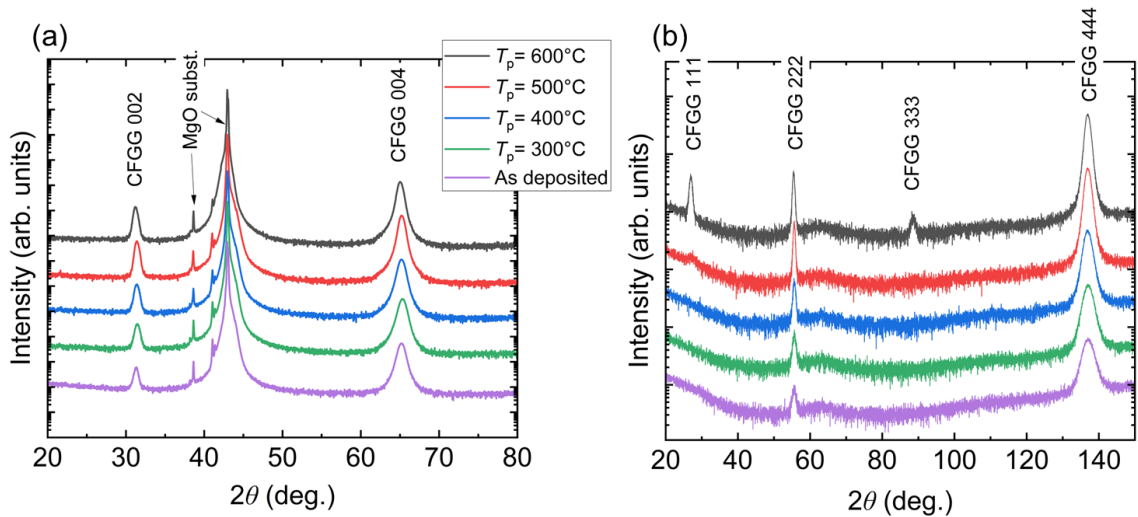

**Fig. 1.:** Out-of-plane XRD profiles at (a) $\chi = 0°$ and (b) $\chi = 54.7°$ for the series-A CFGG thin films with different annealing temperature of RT, 300, 400, 500, and 600 °C. The legends are common for (a) and (b) and data are offset for clarity.

**Damping parameter**

In-plane FMR measurements were performed for the series-A samples to elucidate the effect of $T_p$ on the $\alpha$ of the CFGG films. Supplementary figure S2(a) shows the FMR spectra recorded for the 500 °C annealed CFGG control sample. Figure 2(a) shows the $f$ dependence of $\mu_0 \Delta H$ data fitted using eq. (2) to extract the $\alpha$.

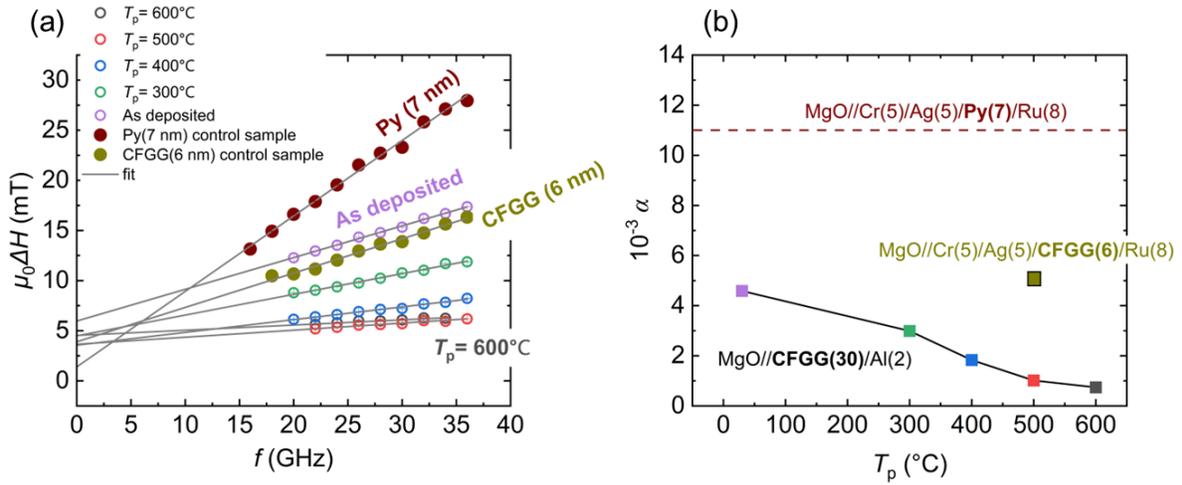

**Fig. 2.:** (a) FMR linewidth ($\Delta H$) vs. frequency ($f$) data for the series-A samples, fitted using equation (2). Symbols represent the experimental data and lines are the fit. (b) Change in damping parameter ($\alpha$) with $T_p$ in the series-A samples, represented by solid square symbols. The $\alpha$ for the 500 °C annealed CFGG control sample is shown by dark yellow square block and the $\alpha$ for the Py control sample is represented by a dotted wine color line.

Figure 2(b) shows the variation of $\alpha$ for the series-A samples. For the 30 nm CFGG films, $\alpha$ decreased monotonically from 0.00459 in the as deposited sample to 0.00074 in the $T_p$ = 600 °C sample. The reduction in $\alpha$ with increasing $T_p$ can be attributed to enhanced atomic ordering[24,38], as indicated by the XRD measurements. To evaluate the effect of $\alpha$ of the FL on the critical current required for magnetization reversal against an applied magnetic field, $\alpha$ was measured for the control samples having structure similar to the FL in the series-B CPP-

GMR stacks. The CFGG control sample with 6 nm thickness exhibited higher $\alpha$ compared to the 30 nm thick CFGG film annealed at the same $T_\text{p}$. Generally, $\alpha$ increases with decreasing thickness due to enhanced interfacial effects[42], including spin-pumping from the CFGG layer to the Ru capping layer[43,44]. However, the $\alpha$ value for CFGG control sample (0.005) is approximately 60%. lower than that of Py control sample (~ 0.011). This lower damping in CFGG is critical for reducing the critical current of the device operation, as discussed later.

**II. Series-B and -C CPP-GMR stacks**

**Structural characterization**

Figure 3 (a) shows the CPP-GMR stack structure for the series-B and series-C samples. Figure 3(b) shows the out-of-plane XRD pattern measured at $\chi = 0°$ for the series-B samples. The observed diffraction peaks corresponding to the (001) plane of Cr, Ag, and CFGG, indicates the (001)-oriented growth. The presence of a superlattice 002 peak from the CFGG layer confirms $B2$ ordering. The diffraction peak corresponding to the Py (220) plane was observed in the Py-FL sample. Figure 3(c) shows the XRD pattern measured at $\chi = 54.7°$, where a 111 superlattice diffraction peak is observed, indicating the presence of $L2_1$ ordering in the CFGG layer. Figure 3(d) and (e) shows the XRD profiles at $\chi = 0°$ and $\chi = 54.7°$ for the series-C samples, respectively. Similar to the series-B samples, (001)-oriented growth and $B2$ ordering of the CFGG layer was observed. The intensity of the CFGG 002 peak increases with the $T_p$, indicating improved $B2$ ordering. Note that the CoFe 002 peak appears close to the Cr 002 and CFGG 004 peaks, making it difficult to distinguish. The CFGG 002 intensity is lower than that in the series-B samples due to the reduced CFGG thickness. The CFGG 111 superlattice peak is not observed for $T_p = 350 - 450$ °C and appears weakly at $T_p = 500$ °C. PendellöUsung fringes are observed in the series-C samples as well as series-B CFGG-FL sample, typically seen in superlattice structures, indicating good interface quality and film uniformity.

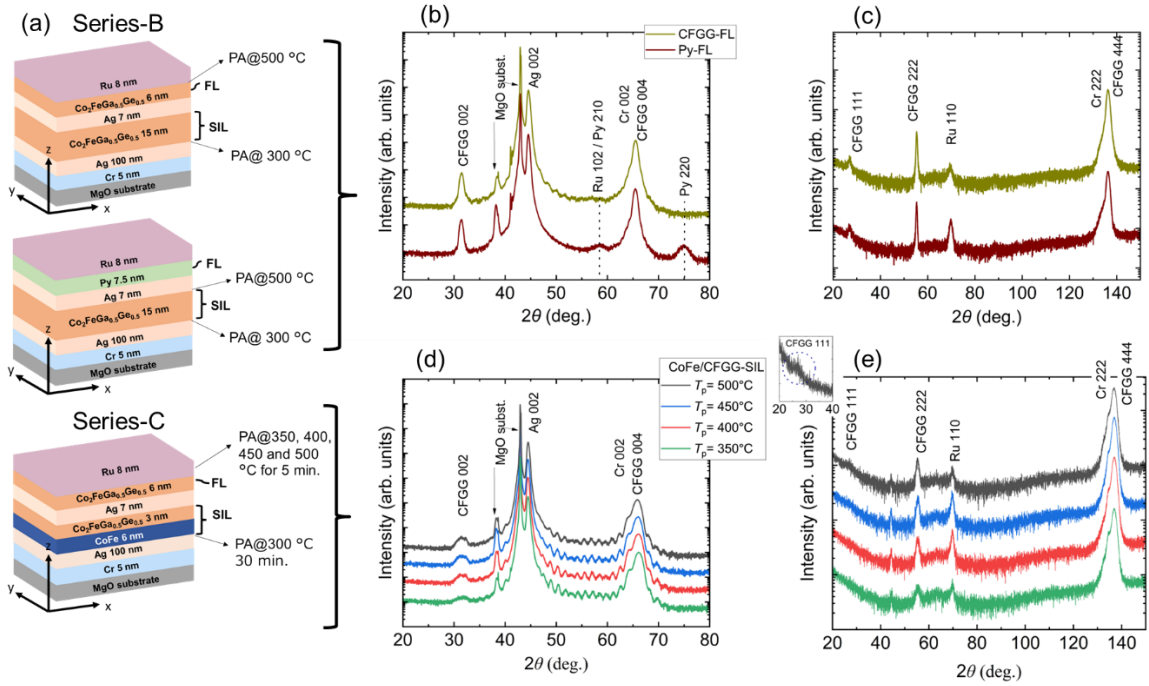

**Fig. 3.:** (a) CPP-GMR stack structures for the series-B and series-C samples. XRD profiles at (b) $\chi = 0°$ and (c) $\chi = 54.7°$ for the series-B samples. XRD profiles at (d) $\chi = 0°$ and (e) $\chi = 54.7°$ for the series-C samples. Inset in (e) shows the zoomed image of the XRD scan for the CoFe/CFGG-SIL ($T_p = 500°C$) sample around the CFGG 111 peak. The legends are common for (b) and (c), and for (d) and (e), with data offset for clarity.

**Magnetotransport measurement (I): MR Measurement**

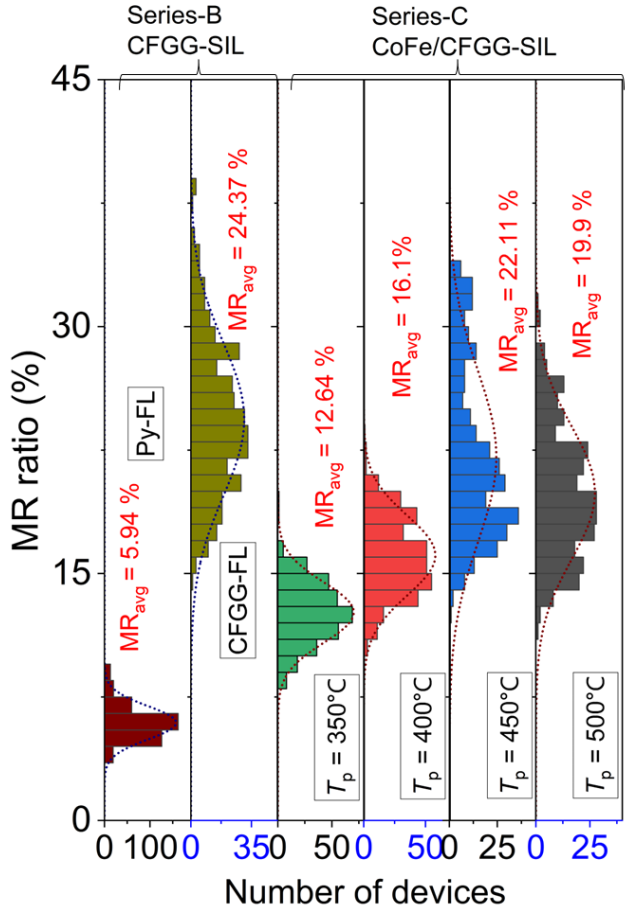

**Fig. 4.:** Distribution of MR ratio in the CPP-GMR devices for the series-B and series-C samples represented by coloured histograms. The abscissa represents the number of devices in each MR ratio interval, as shown on the ordinate axis. Inset in each histogram displays the average MR ratio ($MR_{avg}$).

$R\text{-}H$ measurements were performed on the CPP-GMR devices from the series-B and series-C samples to evaluate their MR properties. Representative $R\text{-}H$ curves for the CPP-GMR devices from both the series are shown in the supplementary figure S3, where the AP configuration of FL and SIL is stabilized by the dipolar interaction at zero magnetic field.

Figure 4 shows the MR ratio distribution for the devices. Overall, the CPP-GMR devices incorporating the CFGG layer in both the SIL and the FL exhibited significantly higher average MR ratio ($MR_{avg}$) ratio compared to the Py-FL sample. This indicates that CFGG is more advantageous for achieving large MR due to its high $P$ and the good electronic band matching at the interface with the Ag spacer[27,28]. The highest $MR_{avg}$ of 24.37 % was recorded for the series-B CFGG-FL sample. In the series-C samples, the $MR_{avg}$ increased with $T_p$, reaching a peak of 21.96% at $T_p$ = 450 °C, followed by a slight decline at $T_p$ = 500 °C. This suggests that 450 °C is the optimal annealing temperature for the CoFe/CFGG-SIL devices. Additionally, the distribution of MR ratios broadened as the $MR_{avg}$ increased, which may be attributed to device-to-device inhomogeneities in the atomic ordering of the CFGG Heusler[45,46].

**Magnetotransport measurement (II): STT Measurement**

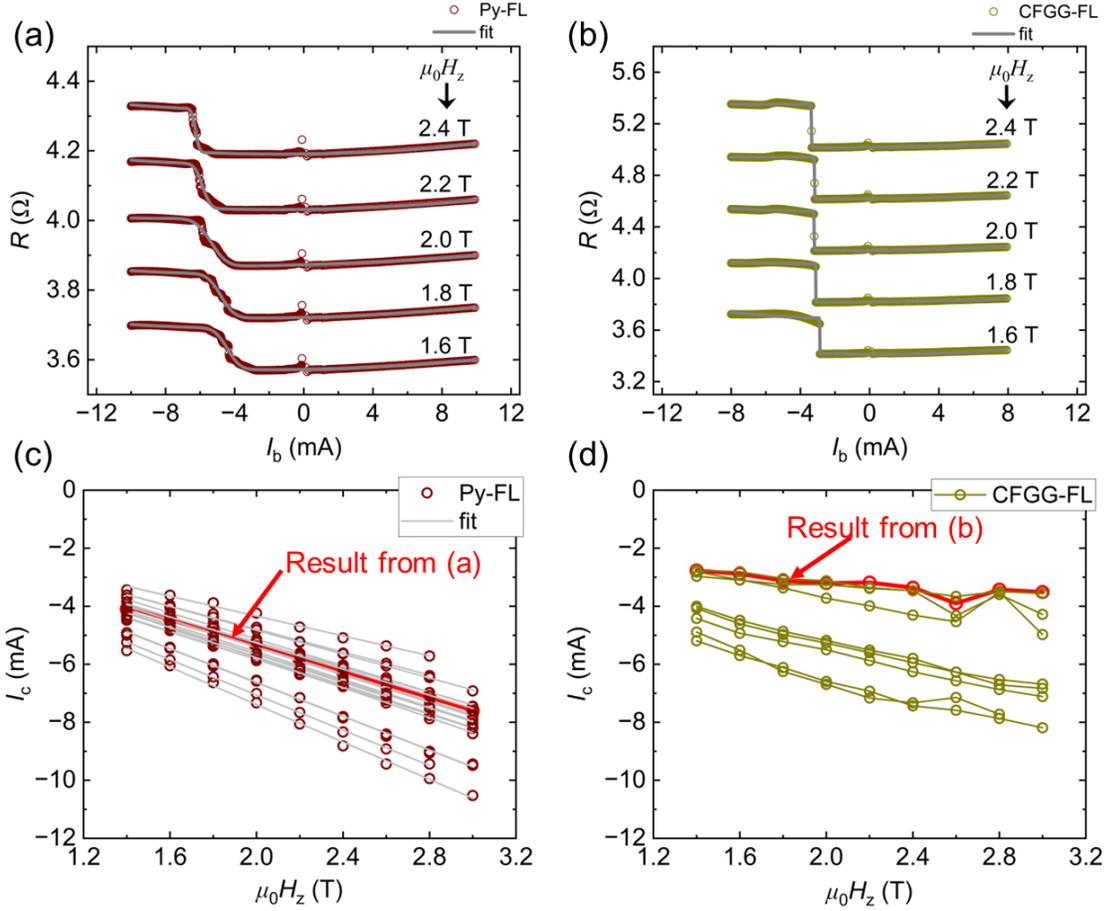

**Fig. 5.:** Exemplary $R$-$I_b$ curves at several constant $\mu_0 H_z$ values for the CPP-GMR device from the series-B (a) Py-FL and (b) CFGG-FL samples. The curves are offset for clarity. Variation of critical current ($I_c$) with $\mu_0 H_z$ for several devices from the series-B (c) Py-FL and (d) CFGG-FL samples. Note: All STT measurements were conducted on devices with standardized lateral dimensions (80 × 80 nm²). As the element sizes are identical, $I_c$ are directly comparable and are used in place of current density.

We investigated the impact of low damping and high spin polarization of CFGG on STT-efficiency by measuring STT-induced FL magnetization reversal against the magnetic field. Figures 5(a) and 5(b) show the exemplary resistance versus bias current ($R$-$I_b$) curves for the series-B Py-FL and CFGG-FL samples, respectively, measured at several constant perpendicular magnetic fields ($\mu_0 H_z$). The $\mu_0 H_z$ values were sufficiently high to align the magnetization of both the FL and SIL parallel to the $\mu_0H_z$ direction at zero $I_b$. The parabolic increase in resistance observed for both positive and negative biases in the $R$-$I_b$ curves is attributed to Joule heating. At negative bias, where electron flows from the FL to the SIL, an additional step in the device resistance is observed when a sufficiently large negative $I_b$ is applied, and this step corresponds to the magnetization reversal of FL detected through the MR effect. The sign of $I_b$ required for the reversal agrees with the expected behavior of STT-induced FL reversal. Additionally, the magnitude of the bias current necessary to induce the magnetization reversal increases with $\mu_0H_z$ because higher $H_z$ requires stronger STT for the FL magnetization reversal. Notably, the magnetization reversal behavior differs between the Py-FL and CFGG-FL samples. The Py-FL sample exhibits gradual resistance change with $I_b$, indicating gradual magnetization reversal from P to AP configuration. In contrast, The CFGG-FL sample exhibits abrupt resistance change indicating abrupt FL reversal. This difference in the behavior of FL magnetization is attributed to the dependence of STT on the relative angle between the FL and SIL magnetizations. In the case of the Py-FL sample, the FL reversal evolves with $I_b$ because the required STT increases with the evolution of the FL magnetization reversal. In the case of the CFGG-FL sample, the angle dependence of STT is large because CFGG with high $P$ is used in both SIL and FL. Therefore, once the magnetization reversal starts, STT increases with the angle, accelerating the FL reversal, which leads to sharp transition from P to AP configuration.

The $R$-$I_b$ curves were fitted phenomenologically using the following equation:

$$R = f(I_b) + \frac{\Delta R}{2}\left(1 + \mathrm{erfc}\left(\frac{I_b - I_c}{I_{\mathrm{width}}}\right)\right), \quad (3)$$

where $f$ is a second-order polynomial function representing the change in resistance (R) due to temperature, $\Delta R$ represents the change in R due to the magnetization reversal, erfc is the error function, $I_c$ is the critical current corresponding to the $I_b$ value at the center of the R change, and $I_{\mathrm{width}}$ is the $I_b$ width of the R change covering approximately 85% of the R change.

Figures 5(c) and 5(d) show the variation of $I_c$ with $\mu_0H_z$ for several devices from the series-B Py-FL and CFGG-FL samples, respectively. The $I_c$ values exhibit a linear change with $\mu_0H_z$ as predicted by the theory. Although the $I_c$ values exhibited device-to-device distribution similar to the MR property, the CFGG-FL sample overall shows lower $I_c$ than the Py-FL sample, demonstrating that lower $\alpha$ of CFGG can reduce the current required for the magnetization reversal.

The $I_c$ versus $\mu_0 H_z$ data for the Py-FL sample were linearly fitted using equation (1) to determine the STT efficiency, $\eta$. For fitting, the following parameters were used for Py: thickness ($d$) ~ 7 nm, $M_s^{FL}$ ~ 0.9 T and $\alpha$ ~ 0.011. The resulting $\eta$ for CFGG was found to be around ~ 0.6, which is higher than the typical value of ~ 0.4 reported for the conventional CoFe system[41], indicating a high STT efficiency due to high $P$ in the CFGG layer. A comparable value of $\eta$ has been previously reported for the $Co_2Mn_{1-x}Fe_xGe$ ($0 \leq x \leq 0.8$) Heusler alloy in CPP-GMR devices[47]. The CFGG-FL sample data were not analyzed by this model as it considers the case where magnetization reversal evolves gradually with the bias current and is not applicable to abrupt magnetization reversal observed in the CFGG-FL sample. The $R$-$I_b$ curves for the CFGG-FL sample were fitted solely to extract the $I_c$ values for the comparison with those of the Py-FL sample.

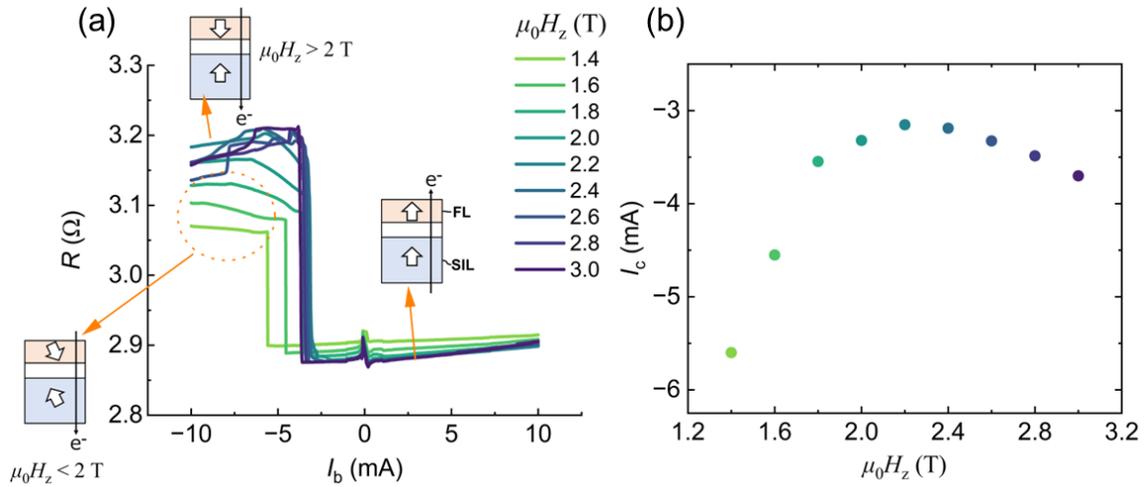

**Fig. 6.:** (a) $R$-$I_b$ curves at several constant $\mu_0 H_z$ values and (b) variation of critical current ($I_c$) with $\mu_0 H_z$ for the MR device from the CoFe/CFGG ($T_p$ = 500 °C) series-C sample. Schematics in the inset of (a) shows the P, AP and incomplete-AP state configuration for the FL and SIL layer magnetizations.

Figure 6(a) shows the exemplary $R$-$I_b$ curves measured at various constant $\mu_0 H_z$ ranging from 1.4 T to 3 T in a CPP-GMR device from the CoFe/CFGG-SIL (annealed at $T_p$ = 500 °C) series-C sample. Resistance steps appear in the negative $I_b$, indicating STT-induced magnetization reversal in the device. The $R$-$I_b$ curves were fitted using equation (3) to extract $I_c$. Figure 6(b) shows the variation of $I_c$ as a function of $\mu_0H_z$ for a representative device. At fields greater than 2.2 T, the magnitude of $I_c$ increases with $\mu_0H_z$ similar to the results from the series-B samples. However, at lower $\mu_0H_z$ values less than 2.2T, the $I_c$ magnitude decreases with $\mu_0H_z$. Furthermore, the resistance steps in Fig. 6(a) are smaller in this low field ($\mu_0H_z < 2.2$T) range compared to those at higher field range ($\mu_0H_z > 2.2$T), suggesting that a complete AP configuration state was not achieved, even under high bias current. This difference between the lower and higher $\mu_0H_z$ range is explained as follows. At lower $\mu_0H_z$, the magnetization of the SIL (CoFe/CFGG bilayer) may not be fully aligned with the applied perpendicular field because of the larger demagnetization field associated with the high magnetization CoFe layer. The resulting tilt in the SIL magnetization reduces the STT efficiency. Additionally, the SIL magnetization might be unstable under counter STT from the spin injection by the FL, which also prevents the establishment of a complete AP configuration. By applying a sufficiently strong $\mu_0H_z$, the SIL magnetization aligns perpendicularly and stabilizes, enabling only the FL magnetization to reverse.

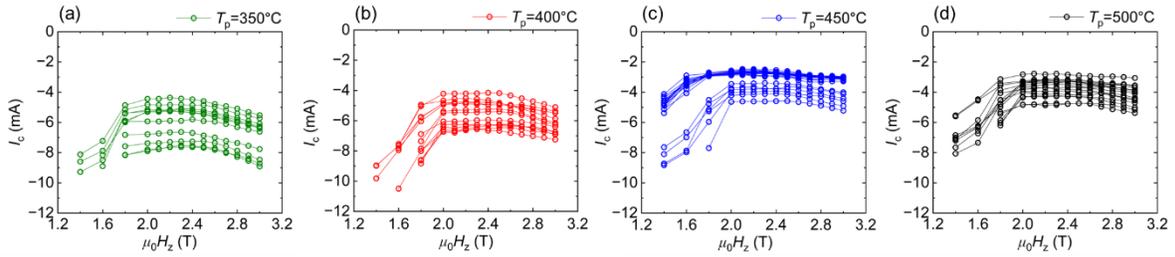

**Fig. 7.:** Variation of critical current ($I_c$) with $\mu_0H_z$ for several MR devices from the series-C samples at different annealing temperatures: (a) $T_p = 350°$C, (b) $T_p = 400°$C, (c) $T_p = 450°$C and (d) $T_p = 500°$C.

Figures 7(a)–7(d) show the variation of $I_c$ values with $\mu_0H_z$ for the CPP-GMR devices from the series-C samples, annealed at different temperatures: 350°C, 400°C, 450°C and 500°C, respectively. Across all samples, the $I_c$ magnitude exhibits a consistent trend: initially decreasing with increasing $\mu_0H_z$ up to approximately 2.2 T, followed by a subsequent linear increase as $\mu_0H_z$ is further increased. A distribution in $I_c$ values was observed, which correlates with the MR ratio of the devices, as discussed below.

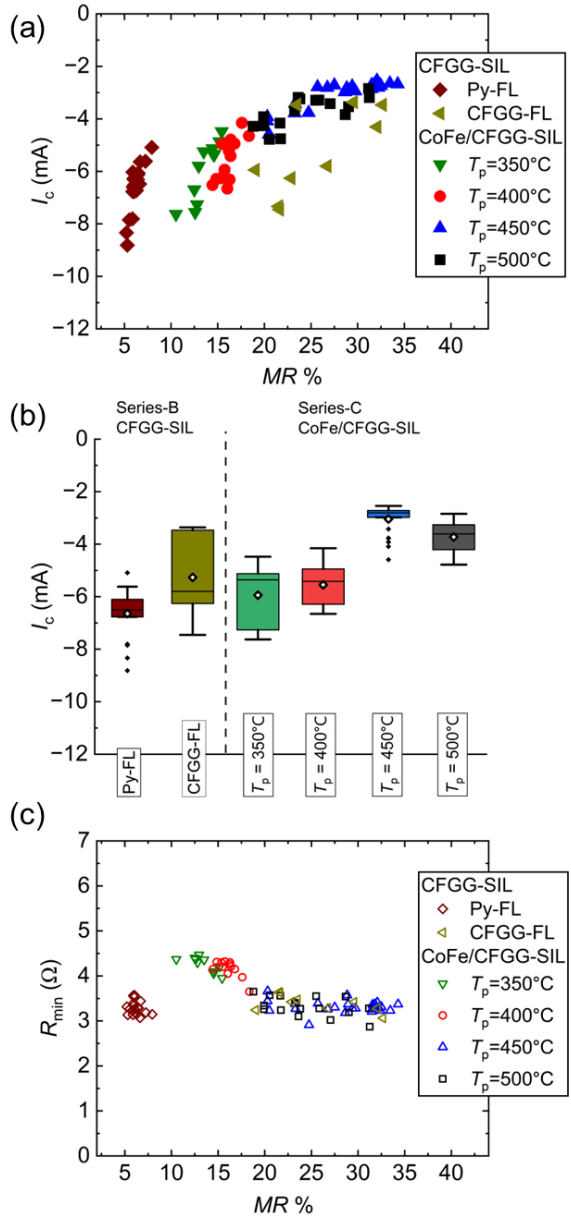

**Fig. 8.:** (a) Critical current ($I_c$) at $\mu_0 H_z$ = 2.4 T plotted against the MR ratios for the measured devices from the series-B and series-C samples. (b) Box plots displaying the distribution of $I_c$ for the same data set shown in panel (a), highlighting statistical variation across different samples, (c) resistance in the parallel state ($R_{min}$) plotted against the MR ratio for the devices corresponding to the data in panel (a).

All STT data presented correspond to devices patterned with standardized 80 × 80 nm² lateral dimensions.

To better understand the distribution of $I_c$ and its correlation with the MR ratio, the $I_c$ values at $\mu_0 H_z = 2.4$ T were plotted against the MR ratios, as shown in Fig. 8(a). We first focus on the results from the series-B samples. As previously discussed, the CFGG-FL sample exhibit lower $I_c$ magnitude than the Py-FL sample due to low $\alpha$ of CFGG as compared to Py. When we compare the best devices from the CFGG-FL and Py-FL samples, the $I_c$ magnitude decreased from -5.1 mA to -3.4 mA, representing a ~ 33% Reduction. Based on the SEM-estimated pillar area, the corresponding current density dropped from $4.8 \times 10^7$ A/cm² to $3.2 \times 10^7$ A/cm². This reduction in the $I_c$ magnitude is smaller than the 60% decrease in $\alpha$ from Py to CFGG, despite that $I_c$ is considered proportional to $\alpha$. One reason for this discrepancy might be the different reversal behavior of the FL magnetization. In the Py-FL sample, the FL magnetization is in the middle of the magnetization reversal at $I_c$, whereas, in the CFGG-FL sample, the FL is almost completely reversed at $I_c$ because of the abrupt transition from P to AP configuration. This difference may hinder the correct comparison of $I_c$ between the Py-FL and CFGG-FL samples and underestimate the reduction in the $I_c$ magnitude for the CFGG-FL sample. It is important to note that the penetration depth of the transverse spin current, over which the transfer of spin angular momentum occurs, is typically less than 2 nm in ferromagnets[48]. The thicknesses of the Py-FL and CFGG-FL are sufficiently higher than this typical transverse spin diffusion length, meaning that the angular momentum of the injected transverse spins is fully transferred to the magnetization. Therefore, the thickness difference in Py-FL and CFGG-FL should not influence the analysis on STT.

Both CFGG-FL and Py-FL samples show an overall positive correlation between the $I_c$ and MR ratio in the device-to-device distributions. This trend is attributed to the fact that devices with high-quality CFGG-SIL layers are expected to show large MR and high $\eta$ simultaneously because of high $P$, leading to reduction in the $I_c$ magnitude. Additionally, in the CFGG-FL samples, devices with high-quality CFGG-FL are expected to have low $\alpha$ in the FL as low $\alpha$ is related to high $P$ of CFGG, influencing the correlation between the $I_c$ and MR.

For the series-C CoFe/CFGG-SIL samples, the $I_c$ magnitude decreased with an increase in $T_p$ from 350 °C to

450 °C, followed by a slight increase at 500 °C. This trend is consistent with that of the MR ratio and is attributed to the increased $P$ and reduced $\alpha$ of the CFGG layers due to the enhanced atomic ordering. The sample annealed at $T_\mathrm{p}$ = 450 °C exhibited the lowest $I_\mathrm{c}$ of approximately -2.5 mA ($J_\mathrm{c} \sim 2.4 \times 10^7$ A/cm²) at $\mu_0 H_z$ = 2.4 T. As with the series-B samples, the device-to-device distributions in the $I_\mathrm{c}$ and MR ratio in the series-C samples showed a general positive correlation, for the same reason discussed above.

Figure 8(b) shows the box plots for the $I_\mathrm{c}$ distributions in the series-B and series-C samples. Each box represents the interquartile range of 25%–75% of the $I_\mathrm{c}$ values for the corresponding sample, with the median indicated as a horizontal line within the box. The white diamond in each box denotes the mean $I_\mathrm{c}$. Whiskers extend to 1.5 times the interquartile range, representing the $I_\mathrm{c}$ range within which most data points fall. Outliers are represented as dots beyond the whiskers. Remarkably, the CoFe/CFGG-SIL ($T_\mathrm{p}$ = 450 °C) sample exhibit a very narrow $I_\mathrm{c}$ distribution, in comparison to the other CoFe/CFGG-SIL samples with different $T_\mathrm{p}$ and the CFGG-FL sample. This suppression in the $I_\mathrm{c}$ distribution is interesting considering that the MR ratio distribution is similarly broad in both CoFe/CFGG-SIL ($T_\mathrm{p}$ = 450 °C) and CFGG-FL sample. These results demonstrate that CoFe/CFGG-SIL is effective in realizing the stable STT, leading to reliable operation in the STT-based devices. One possible reason for this stable STT is that CoFe/CFGG SIL is more robust to counter spin injection than single-layer CFGG SIL because CoFe has higher $\alpha$ than CFGG. The magnetization instability of single-layer CFGG SIL affects spin injection, leading to the distribution in $I_\mathrm{c}$. This speculation is supported by the comparison between the Py-FL and CFGG-FL samples, where distribution in the $I_\mathrm{c}$ versus MR trend is larger and non-monotonic in the CFGG-FL sample because the counter spin injection is stronger in the CFGG-FL sample due to the high $P$ of FL. Another possible reason is that the contributions of the CoFe/CFGG bilayer structure to the effective SIL spin polarization are stable against variations in the atomic ordering of CFGG.

Figure 8(c) shows the $R_\mathrm{min}$ values of the devices plotted against their corresponding MR ratio whose $I_\mathrm{c}$ data are presented in fig. 8(a). Confirming $R_\mathrm{min}$ is necessary to get insight into the device size distribution, as $I_\mathrm{c}$ is sensitive to cross-sectional area. This analysis was performed across multiple devices from different samples, all nominally designed with 80 × 80 nm² lateral dimensions. Note that the MR ratio is less sensitive to the device size as the

MR ratio is normalized by $R_{min}$. The $R_{min}$ values are stable within the multiple devices from the same samples, indicating that the $I_c$ distribution primarily reflects the materials parameters and not of the device size. For the series-C samples annealed at $T_p$ = 350 °C and 400 °C, $R_{min}$ was slightly higher, although the procedure for the device fabrication is the same among the samples. This increase in $R_{min}$ is due to the increased resistivity of the CFGG layer at lower $T_p$ likely originating from the reduced atomic ordering in the CFGG layers, as shown in the supplementary figure S4. Overall, the influence of device size distribution on the presented results can be considered minimal.

**Conclusion**

We investigated the CFGG Heusler alloy in CPP-GMR devices to explore its potential in reducing operational current of STT-induced magnetization reversal, leveraging its high $P$ and low $α$. In comparison between devices using CFGG and Py as the FL, CFGG-based devices demonstrated a 33% reduction in operational current, attributed to its lower α. Furthermore, incorporating a bilayer SIL composed of CoFe and thin CFGG layers led to an additional reduction in operational current, due to spin scattering asymmetry at the CoFe/CFGG interface. The correlation between STT efficiency and MR ratio was thoroughly studied, revealing reduced device-to-device variation in operational current in devices incorporating CoFe/CFGG-SIL structures. These findings demonstrate the advantages of CFGG as a low-damping FL material and CoFe/CFGG as a stable and efficient SIL, establishing CFGG as a promising candidate material for high-performance STT-based spintronic devices.

**Acknowledgements**

This work was supported by the Advanced Storage Research Consortium (ASRC) and MEXT Initiative to Establish Next-generation Novel Integrated Circuits Centers (X-NICS) grant number JPJ011438.

**Data Availability**

The datasets generated during and/or analyzed during the current study are available from the corresponding author on reasonable request.

**Competing interests**

The authors declare no competing interests.

# Supplementary Material

## XRD profiles for the series-A control samples

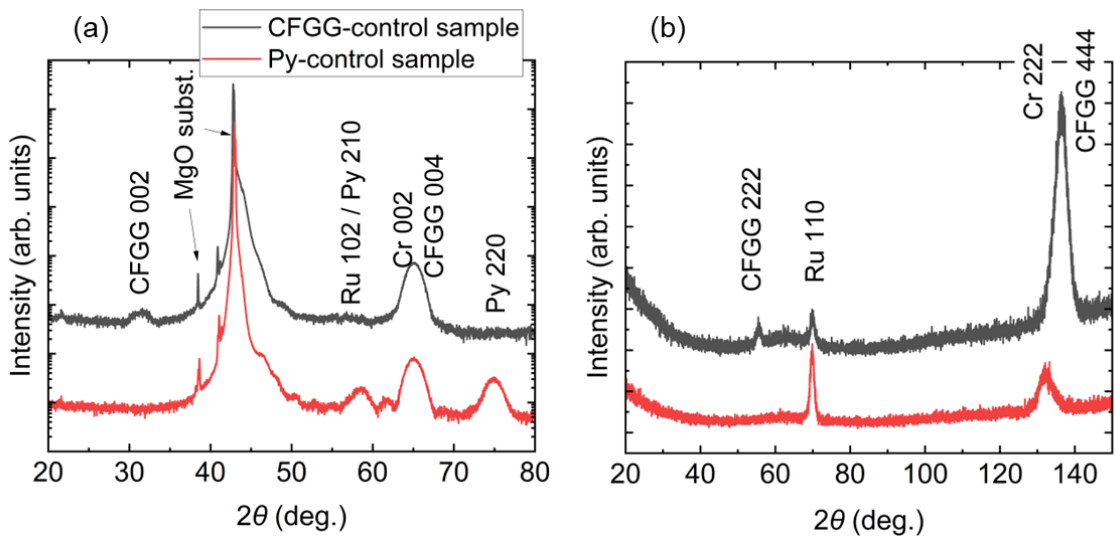

**Fig. S1.:** Out-of-plane XRD profiles at (a) $\chi = 0°$ and (b) $\chi = 54.7°$ for the series-A control samples. The legends are common for (a) and (b) and data are offset for clarity.

Figure S1(a) and S1(b) show the out-of-plane XRD profiles at $\chi = 0°$ and $\chi = 54.7°$, respectively, for the series-A control samples. The XRD profiles at $\chi = 0°$ for the CFGG control sample exhibits diffraction peaks corresponding only to 001 planes, indicating 001-oriented growth. The presence of 002 diffraction peak further confirms *B*2 ordering in the CFGG control sample.

**Ferromagnetic resonance measurement**

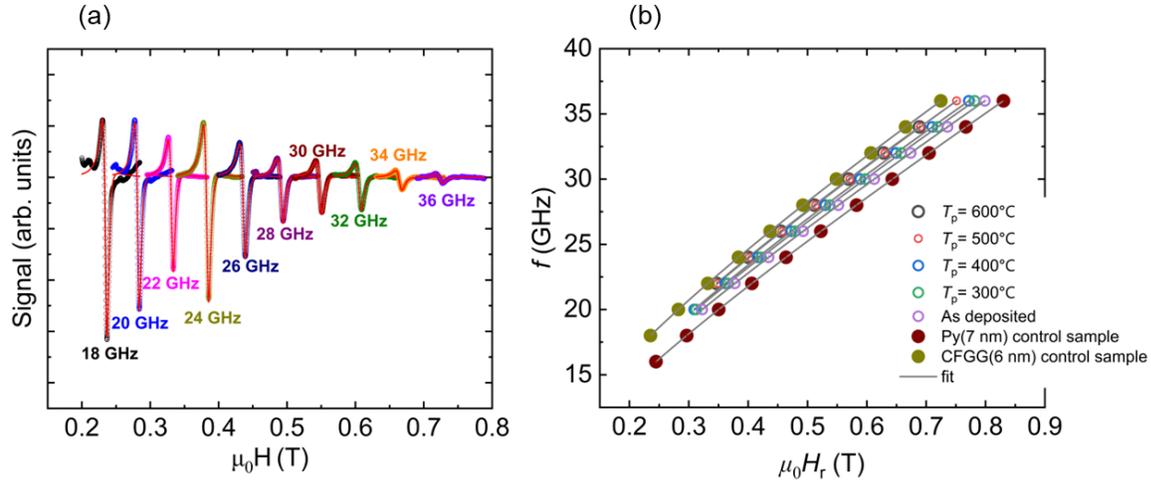

**Fig. S2.:** (a) Exemplary FMR spectra recorded for the 500°C annealed series-A CFGG control sample. Symbols represent the data and lines are the fit to the data. (b) Frequency ($f$) vs. resonance field ($H_r$) data fitted using equation (s1), for the series-A samples. Symbols represent the experimental data and lines are the fit.

Figure S2(a) shows the ferromagnetic resonance (FMR) spectra recorded for the CFGG control sample annealed at 500°C, with the structure Cr(5)/Ag(5)/CFGG(6)/Ru(8).

Figure S2(b) shows the $\mu_0 H_r$ vs $f$ plot, fitted using the Kittel's equation given below

$$f = \frac{\mu_0 \gamma}{2\pi} [(H_r + H_k)(H_r + H_k + M_{\text{eff}})]^{\frac{1}{2}} \qquad (s1)$$

here $\gamma \left(= g \frac{\mu_B}{\hbar}\right)$ is the gyromagnetic ratio with $g \sim 2.1$ as the Lande's factor, $\mu_B$ is the Bohr magneton and $\hbar$ is the reduced Planck's constant, $M_{\text{eff}}$ is the effective saturation magnetization, and $H_K$ is the effective anisotropy field.

**Table S1. Magnetization dynamic parameters for the Series-A thin films.**

Table S1. Magnetization dynamics parameters ($\alpha$, $\mu_0 \Delta H_0$, $\mu_0 M_{eff}$ and $\mu_0 H_K$) obtained from the FMR measurement for the series-A thin films.

| Sample | Line-width broadening, $\mu_0 \Delta H_0$ (mT) | Damping parameter, $\alpha \times 10^{-3}$ | Effective magnetization, $\mu_0 M_{eff}$ (T) | Effective anisotropy, $\mu_0 H_K$ (mT) |
|---|---|---|---|---|
| As deposited | 5.96 ± 0.16 | 4.59 ± 0.08 | 1.0945 ± 0.0055 | 10.40 ± 0.5 |
| $T_p$ = 300 °C | 4.49 ± 0.21 | 2.99 ± 0.11 | 1.1753 ± 0.0049 | 7.70 ± 0.04 |
| $T_p$ = 400 °C | 3.56 ± 0.14 | 1.83 ± 0.07 | 1.2647 ± 0.0028 | -1.28 ± 0.02 |
| $T_p$ = 500 °C | 3.65 ± 0.18 | 1.02 ± 0.09 | 1.3366 ± 0.0115 | -2.06 ± 0.08 |
| $T_p$ = 600 °C | 4.53 ± 0.48 | 0.74 ± 0.07 | 1.3576 ± 0.0086 | -1.98 ± 0.06 |
| Py-control sample | 1.39 ± 0.49 | 11.07 ± 0.27 | 0.9897 ± 0.0093 | -3.94 ± 0.09 |
| CFGG-control sample | 3.9 ± 0.35 | 4.96 ± 0.18 | 1.4061 ± 0.0187 | -0.21 ± 0.09 |

*R-H* curves for the CPP-GMR device from the series-B and series-C samples

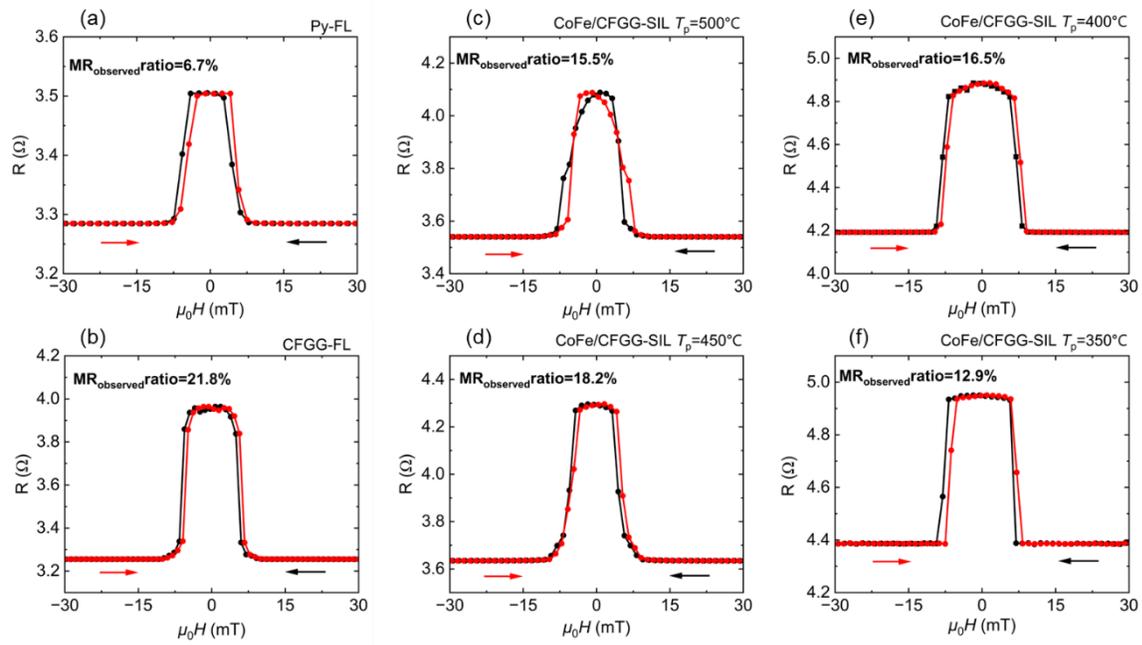

**Fig. S3.:** In-plane *R-H* curves for the selected CPP-GMR devices form the series-B and series-C samples.

Figure S3 shows the in-plane *R-H* curves for the CPP-GMR devices from the series-B and series-C samples.

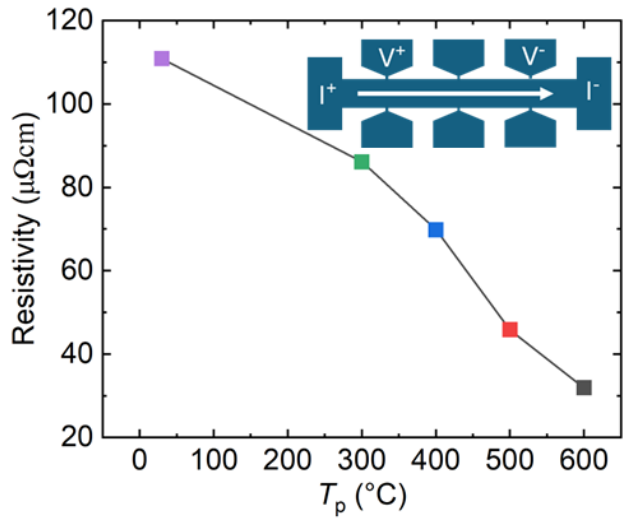

**Fig. S4.:** Change in longitudinal resistivity with annealing temperature ($T_p$) for the series-A samples.

Figure S4 shows the variation in longitudinal resistivity with annealing temperature for the Series-A samples. The samples were patterned into rectangular bars with a length of 2600 μm and a width of 300 μm for resistivity measurements. A gradual decrease in resistivity is observed with increasing annealing temperature, attributed to the improved ordering in the CFGG.